# Coherent excitation-selective spectroscopy in planar metamaterials

Xu Fang,[1] Ming Lun Tseng,[2,3] Din Ping Tsai,[2,3,4] and Nikolay I. Zheludev[1,5]

[1]*Optoelectronics Research Centre and Centre for Photonic Metamaterials, University of Southampton, Southampton SO17 1BJ, UK*

[2]*Graduate Institute of Applied Physics, National Taiwan University, Taipei 106, Taiwan*

[3]*Department of Physics, National Taiwan University, Taipei 106, Taiwan*

[4]*Research Center for Applied Sciences, Academia Sinica, Taipei 115, Taiwan*

[5]*Centre for Disruptive Photonic Technologies, Nanyang Technological University, Singapore 637371, Singapore*

In a proof-of-principle experiment with metamaterials exhibiting electric dipolar and magnetic dipolar resonances, we demonstrated that the electric and magnetic resonances can be separately switches off and on by positioning the metamaterials along a standing wave, while both resonances are present in travelling-wave spectra.

Optical transition between electronic states of matter is governed by selection rules that reflect the symmetry of involved states, which are traditionally defined for travelling electromagnetic waves for spectroscopic application. However, using electromagnetic near-field, the application of the selection rules in absorption spectroscopy could be manipulated as was recently demonstrated for semiconductor nanostructures in the near-field proximity of a noble metal nanoparticle **[1]**, and by placing a quantum emitter in the proximity of metamaterial structures supporting resonances of different symmetry **[2]**.

Here, we show that the optical transitions of different nature in nanoscale thick films of matter can be manipulated by using a standing wave formed by two counter-propagating coherent waves. By placing the thin films at the electric or magnetic antinodes of the standing wave, one can selectively emphasize the electric or the magnetic dipole transitions correspondingly. In a proof-of-principle experiment with periodic arrays of slit nano-antennas and "magnetic wire" metamaterials exhibiting respectively electric dipolar and magnetic dipolar resonances, we demonstrated that the electric and magnetic resonances can be separately switches off and on by positioning the metamaterials along a standing wave, while both resonances are present in travelling-wave spectra. Our demonstration could be generalized to spectroscopic application when a new degree of freedom, the position in the standing wave, is used for the purpose of excitation-selective spectroscopy in sub-wavelength thick films.

The origin of the excitation-selective spectroscopy in the field of a standing wave can be understood by considering $H_{int}$, the Hamiltonian of interaction between electromagnetic radiation and matter, which can written as

$$H_{int} = \frac{1}{c}\hat{d}\frac{\partial \boldsymbol{A}}{\partial t} - \frac{1}{c}(\frac{d\hat{q}^{ij}}{dt} - ce_{ijk}\hat{m}^k)\nabla_j A_i$$

where $\boldsymbol{A}$ is the vector-potential of the electromagnetic field, and $\hat{\boldsymbol{d}}$, $\hat{\boldsymbol{q}}$, and $\hat{\boldsymbol{m}}$ are operators of electric dipole, electric quadrupole and magnetic dipole of the medium, respectively **[3]**. If two linearly polarized coherent counter-propagating waves $A_x = A_0 \cos(\omega t - kz) + A_0 \cos(\omega t + kz)$ oscillating at frequency ω form a standing wave along z-direction, the time dependent Hamiltonian of interaction can be reduced to

$$H_{int} = -2A_0 k [\hat{\boldsymbol{d}}^x \sin(\omega t)\cos(kz) - (\hat{\boldsymbol{m}}^y + \frac{1}{c}\frac{d\hat{q}^{xz}}{dt})\cos(\omega t)\sin(kz)]$$

Hence, if a subwavelength thick layer of matter is placed at the electric nodes of the wave where *cos(kz)=0*, the electric dipole interaction vanishes while magnetic dipole and electric quadrupole interactions become the dominant terms of the Hamiltonian. On the contrary, if the layer is placed at the magnetic nodes where *sin(kz)=0*, the magnetic dipole and the electric quadrupole interactions vanish while the electric dipole interaction becomes the dominant term of the interaction.

We have experimentally demonstrated coherent excitation-selective spectroscopy in two types of metamaterials. In our experiments we used a Ti:sapphire tunable laser to characterize travelling wave and standing wave absorption using an interferometry arrangement [**4**].

In the first case the metamaterial is a planar array of slit nano-antennas (Fig. 1a), the optical response of which is dominated by an electric dipolar resonance. The metamaterial is fabricated from a $Au/Si_3N_4$ film using focused ion beam (FIB) milling (Fig. 1a). The thickness of the whole structure is 80 nm. Fullwave 3D Maxwell simulation spectra of the metamaterial upon travelling-wave excitation are shown in Fig. 1b. An absorption peak at approximately 870 nm is observed, which is associated with the electric resonance arising from the dipolar surface charge oscillation of the nano-antennas [**5**]. Fig, 1(c) shows absorption spectra of the metamaterial in a standing wave. At the electric antinode (E-antinode), the absorption peak at 870 nm is almost twice as strong as that of the travelling-wave. No substantial absorption is seen at the magnetic antinode (H-antinode). Figure 1d presents experimentally measured absorption spectra at the E- and H-antinodes, which qualitatively agree with the theoretical predictions.

In the second case, the metamateraisl is a three-layered structure, the optical response of which is dominated by the magnetic dipolar resonance. The metamaterial is fabricated from a $Au/Si_3N_4/Au$ film using FIB milling (Fig. 2a). The thickness of the whole structure is 110nm. Figure 2b shows the simulated spectra of the metamaterials upon travelling-wave excitation, which indicates a magnetic absorption resonance at approximately 890 nm [**6**]. When the metamaterial is placed at the magnetic antinode of the standing wave, it still exhibits strong absorption at 890 nm. At electric antinode absorption drops below 0.2. Experimental measurement (Fig. 2d) quantitatively confirms these results, in which the absorption maximum can be found at the H-antinode.

Our demonstration could be generalized to spectroscopic application when a new degree of freedom, the position in the standing wave, is used. This excitation-selective spectroscopy will facilitate detection of weak resonances in sub-wavelength thick films.

**Reference:**

Electronic address: x.fang@soton.ac.uk

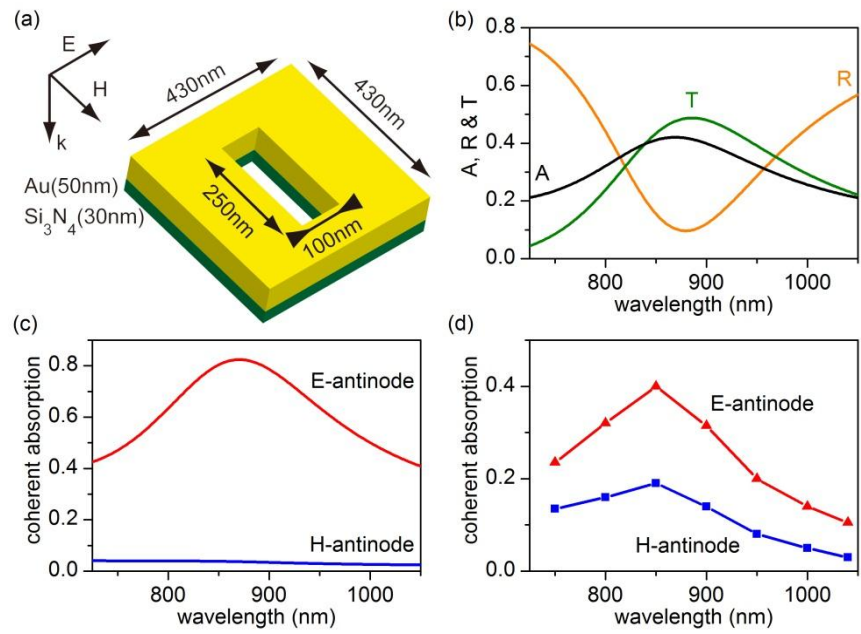

**Figure 1 | Coherent spectroscopy of the slit nanoantenna metamaterial with dominant electric dipolar response.** (a) The unit cell of the metamaterial. (b) Simulated absorption $A$, refection $R$ and transmission $T$ spectra upon travelling-wave excitation. (c) Simulated and (d) measured standing-wave absorption spectra at the electric and magnetic antinodes.

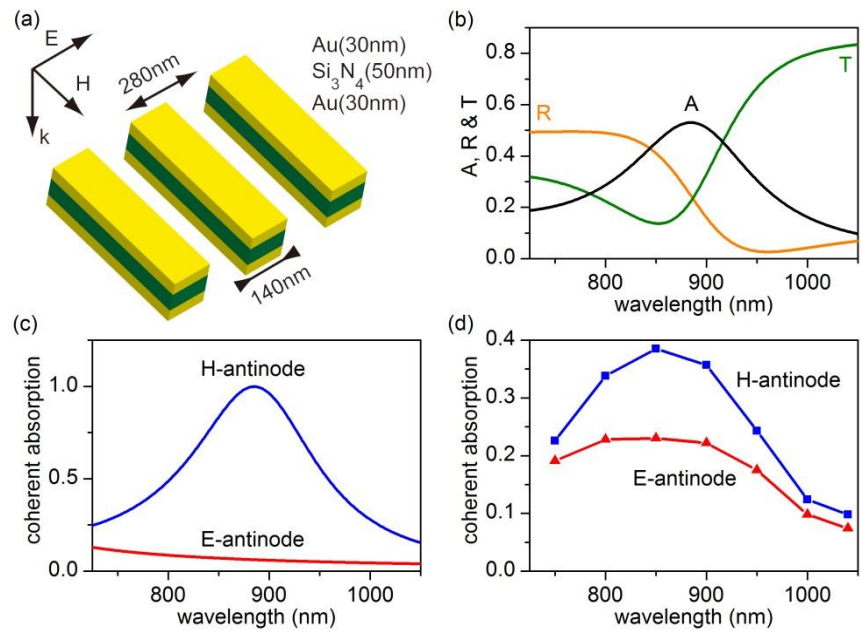

**Figure 2 | Coherent spectroscopy of the "magnetic wire" metamaterial with dominant magnetic dipolar optical response.** (a) The unit cell of the metamaterial. (b) Simulated absorption $A$, refection $R$ and transmission $T$ spectra upon travelling-wave excitation. (c) Simulated and (d) measured standing-wave absorption spectra at the electric and magnetic antinodes.